# DIVISIVE-AGGLOMERATIVE ALGORITHM AND COMPLEXITY OF AUTOMATIC CLASSIFICATION PROBLEMS

Alexander Rubchinsky
National Research University Higher School of Economics, Russian Federation,
International Laboratory of Decision Choice and Analysis
International University "Dubna", Russian Federation,
Department of Applied Mathematics and Informatics,

**Abstract**

An algorithm of solution of the Automatic Classification (AC for brevity) problem is set forth in the paper. In the AC problem, it is required to find one or several partitions, starting with the given pattern matrix or dissimilarity / similarity matrix. The three-level scheme of the algorithm is suggested. At the internal level, the frequency minimax dichotomy algorithm is described. At the intermediate level, this algorithm is repeatedly used at alternations of divisive and agglomerative stages, which causes the construction of a classifications family. At the external level, several runs of the algorithm of the intermediate level are completed; thereafter among all the constructed classifications families the set of all the different classifications is selected. The latest set is taken as a set of all the solutions of the given AC problem. In many cases, this set of solutions can be significantly contracted (sometimes to one classification). The ratio of cardinality of the set of solutions to cardinality of the set of all the classifications found at the external level is taken as a measure of complexity of the initial AC problem.

For classifications of parliament members according to their vote results, the general notion of complexity is interpreted as consistence or rationality of this parliament policy. For "tossing" deputies or / and whole fractions the corresponding clusters become poorly distinguished and partially perplexing that results in relatively high value of complexity of their classifications. By contrast, under consistent policy, deputy's clusters are clearly distinguished and the complexity level is low enough (i.e. in a given parliament the level of consistency, accordance, rationality is high).

The mentioned reasoning was applied to analysis of activity of 2-nd, 3-rd and 4-th RF Duma (Russian parliament,1996-2007). The classifications based upon one-month votes were constructed for every month. Calculation of an average complexity for every Duma have demonstrated its almost three times decrease in the 3-rd Duma as compared to the 2-nd Duma as well as its subsequent essential increase in the 4-th Duma as compared to the 3-nd Duma. The decrease of the suggested index was the most pronounced in 2002 in the wake of the "political peculiar point" – creation of the party "United Russia" 01.12.2001. In 2002 the complexity was equal to 0.096 that was significantly less when in any other year at the consider 12-years period. The introduced notions allow suggesting new meaningful interpretations of activity of various election bodies, including different country parliaments, international organizations and board of large corporations.

# 1. Introduction

An experience in solving of various Automatic Classification (AC) problems, both model and real ones, demonstrates that among them simpler and more complicated problems can occur. In intuitively simple situations finding classifications do not cast any doubt, while in more complicated situations this is not the case. The causes might be different, for instance:

- classifications are not the unique ones;
- the mere existence of classifications is not evident;
- a classification is unique and intuitively clear but it is not clear how it can be found;
- search of classifications in real dimensions leads to significant computational difficulties.

Other reasons can also determine the complexity of AC problems. However, these issues, despite of their practical and theoretical importance, are almost not considered in the literature, except for the analysis of computational complexity of some AC algorithms. Just the absence of the general formal notion of complexity of AC problems, as well as the absence of algorithms of their solutions that cope with problems of various complexity in the framework of one scheme, has initiated the present investigation.

The solution of an AC problem is understood as a family of classifications that includes all reasonable (in some sense) classifications. The complexity of a problem is determined in the construction of the above mentioned family. Generally, the subsequent choice of one or several classifications can be accomplished on a basis of additional data by specialists in the considered specific domain, i.e. beyond the framework of the initial AC problem. The corresponding multi-criteria problem is not considered in the paper; only some reasoning concerning the possible criteria are given. Yet frequently encountered situations, in which intuitively evident solution does exist, are briefly mentioned. Such solutions are selected based on the notions introduced in the paper.

The material is structured as follows. In section 2 the suggested algorithm of the family of classifications construction is briefly described.

Comments, examples and discussion concerning the material of section 2 are presented in section 3. The general formal definition of complexity of an AC problem is introduced in section 4. The results of application of the proposed algorithm for the solution of AC problem and calculation of its complexity to analysis of activity of the 2-nd, the 3-rd and the 4-th RF Dumas (Parliaments) are described in Section 5. In the Conclusion the further possibilities and directions of elaborating of the suggested approach are mentioned.

## 2. Algorithm of solution of AC problem

In this section the algorithm of solution of AC problem is described. As it was mentioned above, all the necessary explications and comments are given in section 3. In the described algorithm initial data about objects' proximity are presented in the well-known form of dissimilarity matrix. This means that all the objects are ordered by indices from 1 to $N$ and for two arbitrary indices $i$ and $j$ numbers $d_{ij}$, interpreted as the degree of dissimilarity or the distance between $i$-th and $j$-th objects, are given. It is assumed that dissimilarity matrix $D = (d_{ij})$ ($i, j = 1, \ldots, N$) is a symmet-rical one; by definition, $d_{ii} = 0$ ($i = 1, \ldots, N$).

Let us give the concise description of the suggested essential algorithm.

At the ***preliminary stage*** the neighborhood graph $G$ is constructed (see subsection 2.1), basing on dissimilarity matrix $D$. At the ***main stage*** both formal objects − neighborhood graph and dissimilarity matrix – are used as inputs.

The algorithm of the main stage is determined as a three-level procedure. At the ***external level*** (subsection 2.4) several runs of the algorithm of the intermediate level are completed. At every run a family of classifications – candidates for solution of the initial AC problem – is determined. Output of the external stage is a new family of classifica-tions, selected among the above mentioned families. This new family is considered as a complete solution of the initial AC problem.

At the ***intermediate level*** one family of classifications is constructed. It is executed by a special ***Divisive-Agglomerative Algorithm*** (***DAA***). DAA description is given in subsection 2.3.

DAA is based on the new algorithm of graph dichotomy (subsection 2.2). It presents the ***internal level*** of the suggested classification algorithm of the general three-level procedure of the main stage.

**2.1. Preliminary stage - neighborhood graph construction.** This notion is well-known (see, for instance, [Luxburg, 2007]). Graph vertices are in one-to-one correspondence to given objects. For every object (say, *a*) all the other vertices are ordered as follows: the distance between *i*-th object in the list and object *a* is a non-decreasing function of index *i*. All the distances are presented in dissimilarity matrix *D*. The first four vertices in this list and all the other vertices (if they exist), whose distance from *a* are equal to the distance from *a* to the 4-th vertex in the list, are connected by edge to the vertex, corresponding to object *a*. It is easy to see that the constructed graph does not depend upon a specific numerations, satisfying the above conditions.

**2.2. Frequency minimax algorithm of graph dichotomy.** The input of the algorithm is an undirected connected graph *G*. There is one integer algorithm parameters: number of repetition *T* for statistics justification.

1. <u>Preliminary stage</u>. Frequencies in all the edges are initialized by 0.
2. <u>Cumulative stage</u>. The operations of steps 2.1 – 2.3 are repeated *T* times:
   2.1. Random choice of a pair of vertices of graph *G*.
   2.2. Construction of a minimal path (connecting the two chosen vertices, whose longest edge is the shortest one among all such paths) by Dijkstra algorithm. The length of an edge is its current frequency.
   2.3. Frequencies modification. 1-s are added to frequencies of all edges belonging to the path found at the previous step 2.2.
3. <u>Final stage</u>.
   3.1. The maximal (after *T* repetitions) value of frequency $f_{max}$ in edges is saved.
   3.2. The operations of steps 2.1 – 2.3 are executed once.
   3.3. The new maximal value of frequency $f_{mod}$ in edges is determined.
   3.4. If $f_{mod} = f_{max}$, go to step 3.2; otherwise, go to the next step 3.5.
   3.5. Deduct one from frequencies in all edges forming the last found path.
   3.6. Remove all the edges, in which frequency is equal to $f_{max}$.
   3.7. Find connectivity components of the modified graph. The component with the maximal number of vertices is declared as the 1-st part of the constructed dichotomy of the initial graph; all the other components form its 2-nd part. After that all the edges, removed at step 3.6, are returned into the graph, except the edges, connecting vertices from different parts of the dichotomy. ∎

Note, that despite the fact of connectivity of the initial graph, the graph presenting the 2-nd part of the dichotomy can be disconnected.

**2.3. Intermediate level – DAA.** This subsection is devoted to DAA description. Its flow-chart is shown in Fig. 1. The neighborhood graph (see subsection 2.1) and dissimilarity matrix together form the input of DAA. Its output will be defined further. The only parameter of DAA is the maximal number $k$ of successive dichotomies. The DAA itself con-sists in alternation of divisive and agglomerative stages.

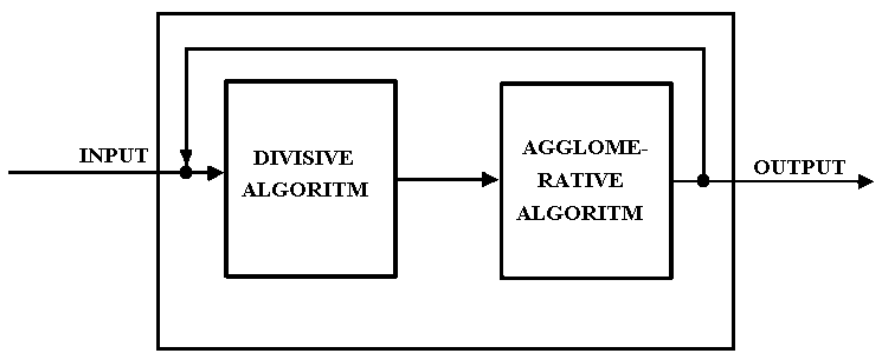

Fig. 1. DAA flow-chart

At the beginning the frequency minimax algorithm of graph dichotomy (see subsection 2.2) divides the initial (neighborhood) graph into 2 parts. Let us denote the found classification into 2 classes as $D_2$. Thereafter one of these two subgraphs, whose number of vertices is larger, is divided by the same algorithm into 2 parts that results in classification $D_3$ of the initial set into 3 classes. Classifications $D_2$ and $D_3$ are named the ***essential*** ones. Denote them as $C_2^2$ and $C_3^3$. After entering the next essential classification $D_j$ ($j \geq 3$) to the agglomerative stage the following operations are completed.

Classification $D_j$ into $j$ classes determines the subfamily of classification into $j$ classes ($D_j$ itself), into $j-1$ classes (obtained by the union of subgraphs, connected by the maximal number of edges), and so on, in correspondence to the convenient agglomeration scheme (successively joining subsets, connected by the maximal number of edges), till to classification into 2 classes. Denote the constructed classifications as $C_j^j$, $C_{j-1}^j$, …, $C_2^j$. These classifications are named the ***adjoint*** ones.

Let us come back to the divisive stage. Among all the classes of the already constructed classification $D_j$ select the class whose graph contains the maximal number of vertices. Check its connectivity. If it is a disconnected one, add one edge connecting two closest vertices belonging to different components. Continue the same operations till the graph becomes a connected one. Just here initial dissimilarity matrix $D$ is used. Completion of these operations guarantees connectivity of graph in the input of the above considered dichotomy algorithm. Applying the frequency dichotomy algorithm to the selected and modified (if necessary) graph, find two new classes. Together with other classes of $D_j$ (except the divided one) these two classes form new essential classification $D_{j+1}$ into $j+1$ classes. Return another time to agglomeration stage and determine adjoint classifications $C_j^{j+1}$, …, $C_2^{j+1}$. Repeating the described steps $k$ times produces the following family of classification:

$$C_2^2; C_2^3, C_3^3; C_2^4, C_3^4, C_4^4; …; C_2^{k+1}, C_3^{k+1}, …, C_{k+1}^{k+1}. \qquad (1)$$

This family is defined as the output of DAA. Pay attention that some classifications from list (1) can coincide to one another.

**2.4. External level – repetitive DAA runs.** At the external level DAA is applied to the same initial graph. However, the output of DAA (list of found classifications) in different runs can differ. The matter is that at every step of accumulating stage a pair of vertices that must be connected by a path is selected randomly. It implies that in AC problems, both model and real, output of DAA depends upon the initialization of random generator. More precisely, some classifications at different DAA runs differ one to another, whereas some classifications coincide at the all DAA runs. Just these distinctions allow us to find "correct" classifica-tions. Therefore it is necessary to complete several runs of the same algorithm with the same initial data − otherwise it is simply impossible to find out in one or another actual situation.

From the formal point of view the situation is clear enough. $r$ DAA runs are executed. The output of this level as well as the final output of the suggested algorithm of solution of AC problem is a family of all the different classifications selected among all the classifications found as a result of $r$ DAA runs. This selection is a standard problem, solved by the direct pairwise comparisons. The possibilities of contraction of this family – sometimes up to one "correct" classification – are discussed in subsection 3.4.

**3. Comment to algorithm of solution of AC program**

**3.1. Frequency minimax algorithm of graph dichotomy.** Let us start with an historical journey. In the article "Community structure in social and biological networks" [Girvan and Newman, 2002] a new approach to graphs decomposition – and thereby to AC problem – was suggested. Let us describe the essence of the matter, citing the article.

"We define the *edge betweenness* of an edge as the number of shortest paths between pairs of vertices that run along it. If there is more than one shortest path between a pair of vertices, each path is given equal weight such that the total weight of all the paths is unity. If a network contains communities or groups that are only loosely connected by a few intergroup edges, then all shortest paths between different communities must go along one of these few edges. Thus, the edges connecting communities will have high edge betweenness. By removing these edges, we separate groups from one another and so reveal the underlying community structure of the graph." The formal algorithm for identifying communities is stated in the article as follows.

Girvan-Newman Algorithm
1. Calculate the betweenness for all edges in the network.
2. Remove the edge with the highest betweenness.
3. Recalculate betweennesses for all edges affected by the removal.
4. Repeat from step 2 until no edges remain.

It is clear that during the execution of the algorithm every increment (by 1) of the number of connectivity components means division of one of groups into two parts, that is an hierarchical structure of groups (or communities) determined only by an initial graph, is obtained as a result. Betweenness calculation is reduced to determination of shortest paths for all pairs of vertices; it is well known that it is a computationally efficient operation with upper estimation $n^2$. Subsequently [Newman, 2004] several modifications of this approach have been suggested, among which the most important are:
- use of random paths (instead of shortest ones) for calculation of edges betweenness;
- use of relatively small part of pairs of vertices (instead of all of them) for estimation of edges betweenness;
- edge removal based on this estimation.

In this connection instead of the notion "edge betweenness" it seems be more convenient to use the notion "edge frequency" keeping in mind a number of an edge inclusions in constructed paths. Taking into account these modifications, an algorithm of graph division into two parts can be described as follows.

<u>Generalized Girvan-Newman Algorithm</u>
1. Set the current frequency at every edge equal to zero.
2. Choose two vertices of the graph.
3. Find by some method a path between vertices chosen at the previous step. If such a path does not exist, go to step 7.
4. Add 1 to frequencies in all the edges included in the path found at step 3.
5. Under certain conditions return to step 2. The example of such conditions is attainment of a large number of execution of steps 2 – 4 or attainment of stochastic stability when the indices of edges with maximal frequency have not been changed for a long time (possibility of different realizations of this step is obvious).
6. Remove an edge with the maximal frequency and return to step 1.
7. Stop. Graph *G* is divided into two or more connectivity components that correspond to the required classes.

It is natural to name the above considered approach as ***the frequency*** one, because it is based on calculation of frequencies of inclusion of graph edges into consecutively constructed paths. It can be applied to every AC problem as soon as it is presented by a graph, particularly, by above mentioned neighborhood graph. The obvious drawback of Girvan-Newman algorithm (outlined by its authors) is that after removal of an edge with the highest betweenness at step 2 all the accumulated statistics about edges betweenness is deleted and, hence, it is not used subsequent-ly. If it has been possible to save these data for consecutive steps, it could essentially accelerate the algorithm. About this issue in the already cited article [Girvan and Newman, 2002] it is written the following. "To try to reduce the running time of the algorithm further, one might be tempted to calculate the betweennesses of all edges only once and then remove them in order of decreasing betweenness. We find however that this strategy does not work well, because if two communities are connected by more than one edge, then there is no guarantee that all of those edges will have high betweenness – we only know that at least one of them will. By recalculating betweennesses after the removal of each edge we ensure that at least one of the remaining edges between two communities will always have a high value." The same is related to the generalized Girvan-Newman algorithm. However, the dichotomy algorithm, described in subsection 2.2, avoids this trap. The essence of the matter is as follows.

In the previously suggested frequency algorithms paths, connecting a next pair of vertices, are traced <u>independently of all the already traced paths</u>. Yet, taking into account all the already traced paths can obtain cuts between two sets of vertices whose all the edges have the <u>same maximal frequency</u>. Then concurrent removal of all the edges with the maximal frequency defines the desired dichotomy of the graph.

It is turned out that before the execution of step 3.6 of the algorithm (see subsection 2.2) the set of all edges whose frequency is equal to the maximal one, indeed contain a cut of graph $G$. There is

**Statement 1**. Before execution of step 3.6:
a) maximal value of frequency over all the edges of the graph is equal to $f_{max}$, where $f_{max}$ is the number, saved at step 3.1;
b) the set of all the edges, whose frequency is equal to $f_{max}$, contains a cut of graph $G$.

**Proof.** Step 3.2 refers to steps 2.1 – 2.3. Finding the next minimax path at step 2, we can encounter one of the following two cases:

 1. There is a minimax path, connecting vertices chosen at step 2.1, whose all the edges have frequencies lesser than $f_{max}$.
 2. Such a path does not exist.

In the 1-st case after every addition 1 (at step 2.3) to frequencies in all the edges of the given path their maximal value does not exceed $f_{max}$. On the other hand, at least in one edge its frequency increases by 1 and at the same time frequency cannot decrease in any edge. Together it means that after some number $t$ of executions of steps 3.2→3.3→3.4→3.2 at step 2.2 we encounter case 2. At case 2 at any path connecting vertices chosen at step 2.1, there is at least one edge, whose frequency is not lesser than $f_{max}$. Because up to now we have encountered only case 1, then, as it was established, all the frequencies do not exceed $f_{max}$. Therefore at any path connecting vertices, chosen at step 2.1, there is an edge, whose frequency is equal to $f_{max}$. Hence, the set of all the edges whose frequency is equal to $f_{max}$, contains a cut of graph $G$. Addition 1 to frequencies in all the edges of the constructed path at step 2.3 and deduction the same edges at step 3.5 does not changes frequencies, that proves a) and b) and, hence, completes the proof of statement 1. ∎

Statement 1 means that in the suggested version of frequency algorithm the necessity of frequency recalculation does not appear. After the only one statistics accumulation the set of edges with maximal value of frequency contains the required cut of the graph.

Figures 2a and 2b demonstrate cases 1 and 2, considered in the proof of Statement 1. The cut itself, of course, depends upon selection of pairs of vertices and distribution of frequencies in edges existing just before the execution of step 3.1. That is the reason of the execution of the cumulative stage, taking the most part of the time. As a result of this stage the required cut became stable in the sense that forming it edges cease to depend upon the number *T* of the constructed minimax paths. Yet this cut can depend on the initialization of random generator. The presence (or absence) of a dependence of the cut (and, hence, the corresponding dichotomy) upon the initialization of random generator turns out the important feature of the <u>AC problem itself other than the used classification method</u>.

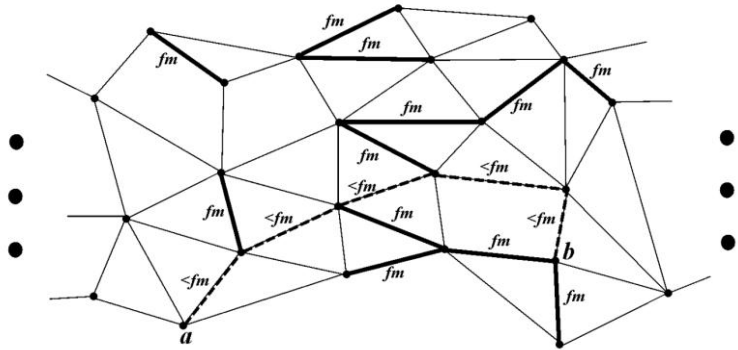

Fig. 2a. Dashed line shows the path, connecting vertices *a* and *b*, in which every edge frequency is less than the maximal frequency $f_m$.

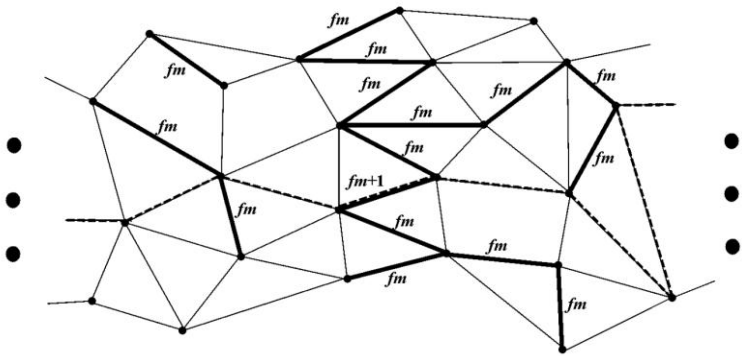

Fig.2b. Dashed line marks the path connecting vertices *b*, located in different sides of the cut, in which all the edges frequency is equal to the maximal one. Such a path compulsory passes along an edge with the maximal frequency $f_m$.

It is also important that in opposite to previously known versions of frequency algorithms, the suggested algorithm finds an approximate solution of some graph optimization problem. This solution expresses a reasonable (even if, like in other cases, incomplete) presentation about correctness of classifications. Let us dwell on it in more detail.

Let us consider connections between the considered algorithm and known optimization statements of a balanced cut in a graph. Introduce the necessary notions. Assume *N* be the number of vertices, *M* be the number of executions of steps 2.1 – 2.3 (but the last one) in the algorithm at stages 2 and 3 together, *A* and *B* be any division of the set of graph vertices, *d*(*A*, *B*) be the cardinality of cut (*A*, *B*). Note that *M* is equal to the number of all the constructed paths in the graph and $M \geq T$.

Consider all the paths (among the constructed ones) whose one end belongs to *A*, and the other end − to *B*. Then sum *S*(*A*, *B*) of frequencies in all the edges from cut (*A*, *B*) is <u>not less than the number of all such paths</u> (denoted as *M*(*A*, *B*)). Indeed, every path increases sum of frequencies at least by one (one, if it intersects cut (*A*, *B*) once, whereas some paths can intersect it several times). Because vertices are chosen at random, probability of the fact that one end of a path belongs to *A* and another to *B* is approximately equal to $(2 \bullet |A| \bullet |B|)/N^2$. Therefore for the total number of such paths there is an approximate equality

$$M(A, B) \approx ((2 \bullet |A| \bullet |B|)/N^2) * M. \qquad (2)$$

Assume (for a rough estimation) that any path from *A* to *B* intersects cut (*A*, *B*) exactly once. Because the number of paths *M* significantly exceeds the maximal value of initial frequency *f*, the following rough estimation takes place:

$$S(A, B) \approx ((2 \bullet |A| \bullet |B|)/N^2) * M. \qquad (3)$$

Dividing both parts of this approximate equality by the number of edges in the cut (*A*, *B*), we receive

$$\bar{f}(A, B) = S(A, B)/d(A, B) \approx (((2 \bullet |A| \bullet |B|)/N^2) * M)/d(A, B), \qquad (4)$$

where $\bar{f}(A, B)$ is the <u>average frequency</u> in edges belonging to cut (*A*, *B*).

It is very important that the suggested algorithm finds cut ($A^*$, $B^*$) whose edges have the same maximal frequency. That means that for any other cut ($A$, $B$)

$$\bar{f}(A, B) \leq \bar{f}(A^*, B^*). \tag{5}$$

Formulae (5) and (4) together mean that cut ($A^*$, $B^*$) maximizes (approximately, in view of made assumptions) expression ((($2 \cdot |A| \cdot |B|$) $/N^2$)*$M$) $/d(A, B)$ over the set of all the cuts of the considered graph. Eliminating from the latest expression constants 2, $N$ and $M$, common for all the cuts, we obtain the expression

$$D(A, B) = \frac{|A| \times |B|}{d(A,B)}. \tag{6}$$

Let us name the function $D(A, B)$ the ***decomposition function*** of a graph. The above reasoning allow to make the following plausible meaningful conclusion: cut ($A^*$, $B^*$), found by the algorithm, approximately maximizes the decomposition function (6) of the considered graph. The fact that in some cases this cut depends upon the initialization of random generator (and for this reason alone it cannot exactly maximize function (6) defining only by the graph itself) just expresses the approximate character of solution of this optimization problem. The corresponding examples are given below in this subsection.

In the above cited review [Luxsburg, 2007] the minimization problem

$$R(A, B) = d(A, B) \times (\frac{1}{|A|} + \frac{1}{|B|}) \to \min, \tag{7}$$

named "Ratio Cut Problem" was considered. Direct comparison of formulae (6) and (7) demonstrates that problems of function $D(A, B)$ maximization and of function $R(A, B)$ minimization (determined on the same set of all cuts of the graph) are equivalent ones. Therefore the suggested frequency algorithm can be used for approximate solution of this well-known "Ratio Cut Problem". Moreover, it is an efficient approximate method for this purpose. Yet the essential question, concerning this *NP*-complete decomposition problem, does not consist in finding its approximate solutions. It rather can be stated as follows: is it true that the ***exact solution*** of the above optimization problem (found by any way) can be considered as an intuitively correct dichotomy? Of course, this question is meaningful and it can be answered only by examples. Several successful examples of correct dichotomies, found by the suggested frequency algorithm, are presented in preprint [Rubchinsky, 2010]. But it is not necessarily the case for arbitrary AC problems.

Just the last circumstance has initiated the elaboration of the general AC algorithm described in this work, in which the suggested algorithm of dichotomy is used as an essential step at the divisive stage (see subsection 2.3). In order to explain the necessity of more thorough analysis the following example is considered.

**Example 1.** Two two-dimensial sets are shown in Fig. 3a and 3c. The dichotomy result for the set of Fig. 3a is shown in Fig. 3b. The cut, found by the frequency algorithm, maximizes the decomposition function (6) over the set of all the cuts of the neighborhood graph and determines intuitively correct classification into two classes. It is reasonable that the same cut minimizes function (7). The result does not depend upon initialization of random generator.

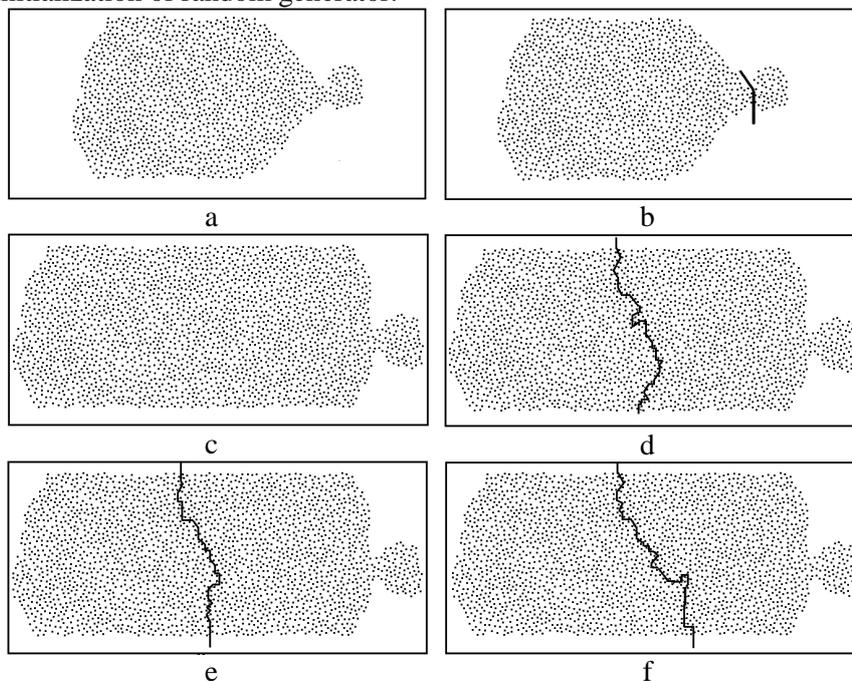

Fig. 3. Examples of found dichotomies. For the dichotomy in Fig. 3b $D = 30758$, in Fig. 3d $D = 40382$, in Fig. 3e $D = 40755$, in Fig. 3f $D = 36886$

At the same time the use of the same algorithm for the similar set, shown in Fig.3c, leads to results, perceptibly depending upon an initialization of random generator, as it is clear from Fig. 3d, 3e, and 3f. In these cases the found solutions do not coincide with intuitively obvious one. Finally, the value of decomposition function for the correct cut is equal to 31549, whereas for the incorrect cut, found by the frequency algorithm and shown in Fig. 3d, it is equal to 40382. In two other cases this function also is essentially greater, than its value on the correct cut. Note, that we are dealing with exact but not approximate values of decomposition function. This simple example another time underlines the caution, which is required in using well-accepted balanced criteria of classification (as well as other formal models of classification).

The cause of failure of criteria (6) in the considered case is clear enough. The ratio between the maximal and the minimal numbers of points, belonging to correct classes, in the set in Fig. 3c is essentially greater than in the set in Fig. 3a. Therefore the numerator $|A|\times|B|$ in (6) is so small relatively to the cardinality of product of approximately equal parts, so that it cannot be compensated by the denominator in (6) equal to relatively small number of edges in the correct cut. The same phenomenon concerns (and even to a greater extent because it is revealed under lesser relation of cardinalities) to other frequency algorithms of dichotomy. ∎

Taking into accounts results of tens computational experiments with different data, we reached the following informal conclusions.

1. The exact solution of the well-known balanced cut problem (and, hence, spectral and kernel methods that approximate this solution) can lead to intuitively wrong classifications in many relatively simple cases.

2. All the stochastically stable dichotomies found by the suggested frequency algorithm are intuitively correct; they maximize criterion (6).

3. All the stochastically unstable dichotomies found by the suggested frequency algorithm are intuitively incorrect; values of criterion (6) exceed its value on the "correct" cut.

Yet the notion of dichotomy stability itself is not the exactly defined one. Between obviously stable and obviously unstable situations there is some "gray zone" of weak instability. Analogously to many situations of such a kind, encountering in various domains of pure and applied mathematics, these intermediate situations in some sense are inevitable, while the most important and intriguing phenomena occur just in such intermediate zones. These reasons do not only initiate but in some sense warrant the suggested approach to AC problems, because it does not only explain but uses in the algorithms instability of classifications.

In summary of this subsection let us note that the only parameter of the frequency algorithm – number *T* of paths at the accumulating stage – is not the essential one. Parameter *T* can be removed, if calculations ceases at reaching stability, i.e. selection of the same cut. If the objects number does not exceed 1000, typical value of repetitions is 1500 – 2000. As noted above, this cut itself can depend upon initialization of random generator, which determines initial values of frequency as well as the sequence of random minimax paths.

**3.2. Intermediate level – DAA.** In order to keep strong properties of the suggested method of dichotomy and to be got rid of its weakness it is natural to consider consecutive dichotomies. For instance, the use of the same algorithm for the maximal (in number of points) of two classes, shown in Fig. 3d, results in classification into three classes, shown in Fig. 4. If now to pool two classes, connected by the largest number of edges, then just the correct classification is obtained. DAA from subsection 2.3 just describes consecutive operations, required to obtaining correct classifications in the general case.

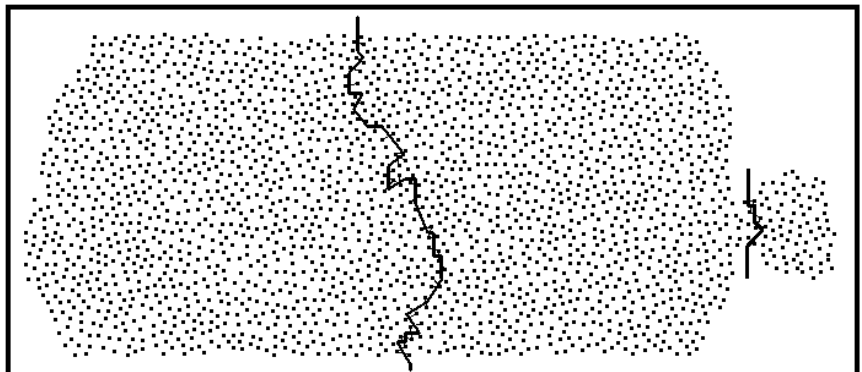

Fig. 4. Result of two consecutive dichotomies

**Example 2.** Let us demonstrate DAA in more complicated case – set of points shown in Fig. 5a. Consider consecutive dichotomies and construction of essential and adjoint classifications, using notation from subsection 2.3. Assume $k = 3$, i.e. restrict our consideration to 3 consecutive dichotomies. Essential classifications $D_2 = C_2^2$, $D_3 = C_3^3$ and $D_4 = C_4^4$ are

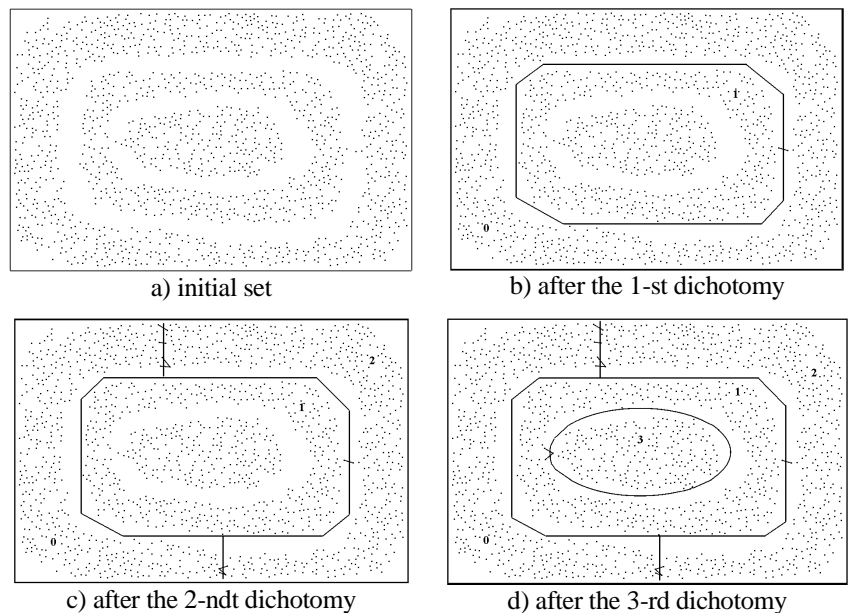

a) initial set      b) after the 1-st dichotomy
c) after the 2-ndt dichotomy      d) after the 3-rd dichotomy
Fig. 5. Initial set and essential classifications

shown in Fig. 5b, 5c и 5d. The edges forming cuts between different classes are shown, too. Pooling classes 0 and 2 from classification $C_3^3$ results in adjoint classification $C_2^3$, coinciding with the essential classification $C_2^2$.

Further, pooling classes 0 and 2 from classification $C_4^4$, shown in Fig. 5d results in adjoint classification $C_3^4$, shown in separate Fig. 6. It is clear that this classification is the desirable "correct" classification. However, DAA does not "know" yet about it and continues the considered agglomerative stage. Pooling classes 0 and 1 from classification $C_4^4$ are connected by 2 edges. Their pooling results in adjoint classification $C_2^4$, coinciding with classifications $C_2^2$ and $C_2^3$.

At this point the work of DAA is over. 6 classifications: $C_2^2$; $C_2^3$, $C_3^3$; $C_2^4$, $C_3^4$, $C_4^4$ are found. Among them there are 4 different classifications: $C_2^2$, $C_3^3$, $C_3^4$, $C_4^4$, shown in Fig. 5b, 5c, 6 and 5d, correspondingly. Pay attention that the correct classification is the adjoint one. It cannot be an essential classification after any number of consecutive dichotomies. It cannot be found as well as a result of agglomerative procedure, starting with one-element or little classes, because rings 1 and 0 + 2 (Fig. 6) cannot be constructed by pooling of closest classes. In DAA just the alternation of divisive and agglomerative stages is especially important. ∎

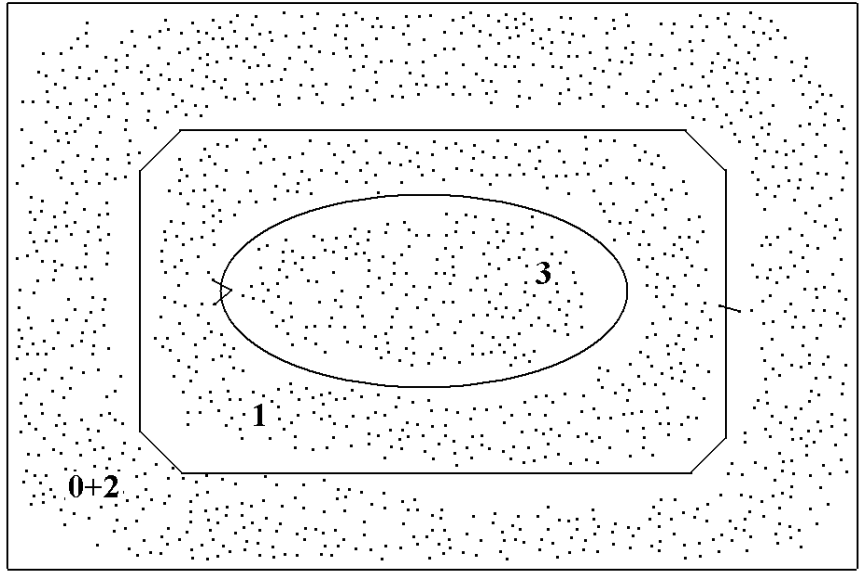

Fig. 6. Correct adjoint classification

**3.3. External level – repetitive DAA runs.** At this stage results of several DAA runs for the same neighborhood graph are considered and compared one to another. Let us consider the encountered situation for the AC problem from example 2.

**Example 3.** Assume (for visibility of illustration) the number of runs $r = 4$. In Fig. 7 results of 4 runs for essential classification $C_3^3$ are shown (see also Fig. 5c). All the 4 found classifications are the different ones.

It is easy to understand that in the same run essential classifications $C_4^4$ are differ of the classifications shown in Fig. 7 only in presence of another class in the center (see also Fig. 5d). This implies that all these four classification also are different ones. At the same time essential classification $C_2^2$ and adjoint classification $C_3^4$ found at all the runs coincide with classifications shown in Fig. 5b and Fig. 6, i.e. they are permanent. ∎

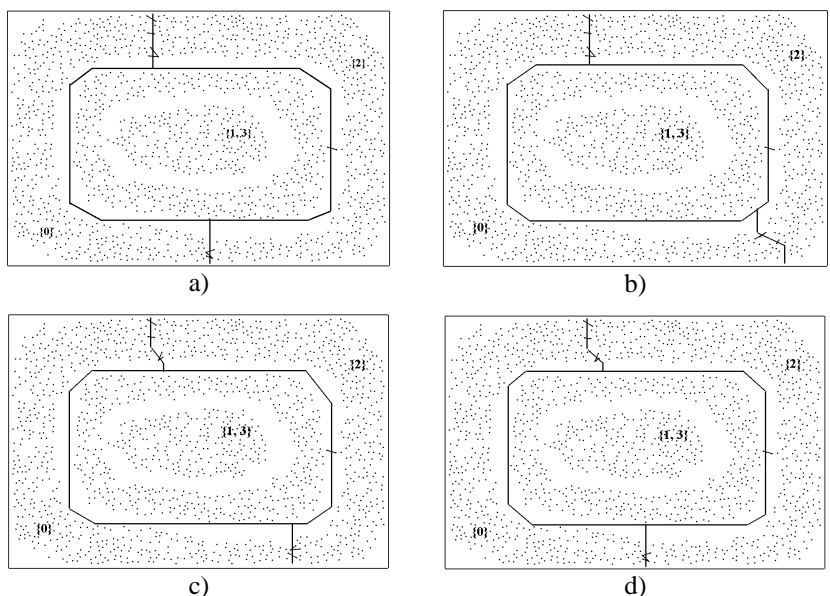

Fig. 7. Classifications $C_3^3$ found at four DAA runs

Thus, the final result, produced by the suggested algorithm, consists of 10 different classifications. Among them there are 8 varying with every run, and 2 permanent classifications.

**3.4. Contraction of classification family.** In many AC problems, partially, in all the model examples considered in preprint [Rubchinsky, 2010], the only correct classification was determined simply enough. A stable (i.e. repeating in all the runs) classification with the maximal number of classes turns out to be the intuitively correct one. In examples 2 and 3 such a classification is shown in Fig. 6. Growth of runs number *r* and dichotomies number *k* nothing changes – no one new stable classification arises, while found classification remains stable. Therefore in such simple situations choice of parameters *r* and *k* can be done adaptively, notifying stable classifications and ceasing calculations, if new stable classifications with greater number of classes do not arise.

Yet in real AC problem the situation proves to be another one. Only "degenerated" classifications are the "absolutely" stable, i.e. repeating completely in all the DAA runs. Classifications are named degenerated if they include one- or two-elements classes. Found meaningful classifications are not absolutely stable: in different runs they coincide, for instance, by 99% but not by 100%.

In order to analyze such situations it is supposed to introduce reasonable criteria, which characterize single classifications. Two criteria are considered as the essential ones: stability and number of classes.

Stability is understood here as a degree of repeatability of a classification under different runs. From the formal point of view the situation is rather simple and well-known. To compare two classifications of the same set RAND index (see, for instance, [Mirkin, 2006, subchapter 7.3]) is used. It is define as follows. Assume $\varphi(i, j) = 1$ iff (if and only if) $i$-th and $j$-th elements are included in one class in both classifications or $i$-th and $j$-th elements are not included in one class in both classifications. In all the other cases $\varphi(i, j) = 0$. Function $\varphi(i, j)$ is summing up over all the pairs of non-coinciding $i$ and $j$; thereafter the sum is divided by the number of all such pairs. The obtained value is equal to 1 iff both classifications completely coincide. This value is named **RAND index** and denoted by $R(A, B)$, where $A$ and $B$ – two classifications of the same set.

Thereafter for any family F of classifications, taken by one from every run, the concordance of the family:
$$c(F) = \min_{A,B \in F} R(A, B). \tag{8}$$
Finally, stability $s(A)$ of classification $A$ is defined as the maximal concordance of family F, contained $A$. Under the introduced definitions calculation of any classification can be executed by computationally efficient greedy algorithm. Stability $s(A)$ of classification $A$ is equal to 1 iff it is completely repeated in all the runs.

The number of classes is a clear criterion, which, of course, does not require any calculations. The other criteria depend upon a specific AC problem.

In many cases the set of all the classifications found by the suggested algorithm can be notably contracted, if among several close (i.e. with pairwise RAND index close to 1) classifications to select only one by the elimination of several degenerated classifications with the greater number of classes. It is expedient to take the number of classes into account after this operation. The alternative approach consists in use of the criterion of uniformity of a classification (ratio between maximal and minimal cardinality of classes in this classification).

The final choice of a single classification among several ones found by the suggested approach, like in other multi-criteria problems, remains to decision-maker.

The material of the present subsection has a preliminary, "sketch" character. The importance of this issue requires the special consideration, including specific examples and conclusions. However, one thing can be stated with certitude. Reasonable solution of real AC problems can be obtained using an interactive computer system, including computational algorithms as well as means of presentation, analysis and visualization of results, which take into account specifics of a considered problem.

## 4. Complexity of AC problems

Analyzing AC problems it is useful to have some objective indices, describing their complexity, entanglement, and other hardly defined properties. These indices must be relevant to arbitrary AC problems rather to its special types.

In the presented work such an index is suggested. It concerns the number of classifications in the set of all the solutions of an AC problem, defined at the end of subsection 2.4. Yet this number depends on the number $k$ of dichotomies in DAA and of number $r$ of DAA runs. It is easy to see that the general number of classifications, considered at the external level of the algorithm, is equal to $\frac{(k+1)*k}{2} * r$. Among them all the different classifications are selected. It seems that a reasonable measure of complexity of an AC problem is ***the ratio between the number of actually existing different classifications and its maximal possible number*** $\frac{(k+1)*k}{2} * r$.

In the AC problem from Examples 2 and 3 for $k = 3$ and $r = 4$ there are 10 different classifications. Dividing 10 to $24 = \frac{4*3}{2} * 4$, we receive 0.417. This is the complexity (in the introduced sense) of the considered AC problem. As in some other domains of discrete mathematics, the introduced notion of complexity of an AC problem is not defined through its initial description but through one specific method of its solution. Therefore the only approach to substantiation of the introduced notion consists in possibility of its meaningful interpretation in actual AC problems.

This issue is considered in the next section.

### 5. Analysis of voting in 2-nd, 3-rd and 4-th RF Duma

In this section activity of State Duma during the period since the beginning of 1996 till the end of 2007 is considered. Many important political events had happened during this 12-year period. And yet, it seems that the separate events were not as important as the process of building of still actual system of political power itself.

Mathematical models of political processes in the first four Duma were considered in detail in the monograph [Aleskerov et all, 2007] and in cited where literature.

For every separate month of the considered period all the votes are considered. To every $i$-th deputy ($i = 1, 2, …, m$) a vector $v_i = (v_1^i, v_2^i, …, v_n^i)$ is related, where $n$ is the number of votes in a given month. Note, that the number $m$ of deputies, though slightly, changed from period to period. Of course, at every moment the number of deputies is always equal to 450. Yet during 4 years some deputies dropped out while the other ones came instead. The number of deputies participated in Duma voting activity in 1996-1997 was equal to 465, in 1998-1999 – to 485, in 2000-2003 – to 479 and in 2004-2007 – to 477.

Assume

$$v_j^i = \begin{cases} 1, & \text{if } i\text{-th deputy voted for } j\text{-th proposition;} \\ -1, & \text{if } i\text{-th deputy voted against } j\text{-th proposition;} \\ 0, & \text{otherwise (abstained or not participated).} \end{cases}$$

Dissimilarity $d_{st}$ between $s$-th and $t$-th deputies is defined as usual Euclidian distance between vectors $v_s$ and $v_t$. The dissimilarity matrix $D = (d_{st})$ is the initial one for finding deputies classes by the method, described in section 2.

The following Tables 1, 2 and 3 present the complexity of corresponding classifications for every month of the voting activity of 2-nd, 3-rd and 4-th RF Duma. The numbers in the 1-st column are the dates (year and month). The numbers in the 2-nd column are equal to the number of votes in the corresponding months. Numbers in the 3-rd columns are equal to complexity of the corresponding AC problem, calculated following the definition of this notion in section 4. Here the number $k$ of consecutive dichotomies is equal to 10, the number $r$ of DAA runs also is equal 10, so that the maximal number $\frac{(k+1)*k}{2} * r$ of classifications is equal to 550. Some reasons, concerning choice of these essential parameters, are discussed

Table 1
**Complexity of voting generated classifications in 2-nd Duma (1996-1999)**

| 1 | 2 | 3 | 1 | 2 | 3 |
|---|---|---|---|---|---|
| 9601 | 174 | 0.610909 | 9801 | 248 | 0.421818 |
| 9602 | 321 | 0.625455 | 9802 | 366 | 0.330909 |
| 9603 | 295 | 0.581818 | 9803 | 347 | 0.469091 |
| 9604 | 470 | 0.683636 | 9804 | 334 | 0.436364 |
| 9605 | 263 | 0.938182 | 9805 | 292 | 0.398182 |
| 9606 | 269 | 0.827273 | 9806 | 489 | 0.534545 |
| 9607 | 450 | 0.263636 | 9807 | 493 | 0.352727 |
| 9608 |  |  | 9808 |  |  |
| 9609 |  |  | 9809 | 405 | 0.390909 |
| 9610 | 432 | 0.494545 | 9810 | 326 | 0.507273 |
| 9611 | 226 | 0.567273 | 9811 | 338 | 0.327273 |
| 9612 | 566 | 0.465455 | 9812 | 534 | 0.392727 |
| 9701 | 234 | 0.456364 | 9901 | 416 | 0.207273 |
| 9702 | 427 | 0.445455 | 9902 | 354 | 0.250909 |
| 9703 | 334 | 0.381818 | 9903 | 482 | 0.369091 |
| 9704 | 437 | 0.316364 | 9904 | 384 | 0.372727 |
| 9705 | 169 | 0.485455 | 9905 | 228 | 0.449091 |
| 9706 | 762 | 0.238182 | 9906 | 768 | 0.392727 |
| 9707 |  |  | 9907 |  |  |
| 9708 |  |  | 9908 |  |  |
| 9709 | 337 | 0.201818 | 9909 | 292 | 0.241818 |
| 9710 | 354 | 0.247273 | 9910 | 338 | 0.270909 |
| 9711 | 253 | 0.289091 | 9911 | 696 | 0.218182 |
| 9712 | 530 | 0.265455 | 9912 | 243 | 0.430909 |

further. The missed rows in Tables 1, 2 and 3 correspond to the months without any voting activity.

Table 2
**Complexity of voting generated classifications in 3-rd Duma (2000-2003)**

| 1 | 2 | 3 | 1 | 2 | 3 |
|---|---|---|---|---|---|
| 0001 | 71 | 0.547273 | 0201 | 279 | 0.183636 |
| 0002 | 228 | 0.112727 | 0202 | 380 | 0.063636 |
| 0003 | 177 | 0.387273 | 0203 | 311 | 0.081818 |
| 0004 | 368 | 0.112727 | 0204 | 640 | 0.114545 |
| 0005 | 279 | 0.141818 | 0205 | 353 | 0.138182 |
| 0006 | 454 | 0.149091 | 0206 | 956 | 0.072727 |
| 0007 | 301 | 0.078182 | 0207 | | |
| 0008 | | | 0208 | | |
| 0009 | 144 | 0.154545 | 0209 | 329 | 0.120000 |
| 0010 | 371 | 0.169091 | 0210 | 541 | 0.067273 |
| 0011 | 240 | 0.103636 | 0211 | 448 | 0.065454 |
| 0012 | 483 | 0.138182 | 0212 | 531 | 0.058182 |
| 0101 | 141 | 0.109091 | 0301 | 144 | 0.203636 |
| 0102 | 254 | 0.245455 | 0302 | 350 | 0.136364 |
| 0103 | 268 | 0.085454 | 0303 | 382 | 0.160000 |
| 0104 | 409 | 0.187273 | 0304 | 519 | 0.136364 |
| 0105 | 248 | 0.296364 | 0305 | 248 | 0.141818 |
| 0106 | 683 | 0.069091 | 0306 | 677 | 0.083636 |
| 0107 | 825 | 0.132727 | 0307 | | |
| 0108 | | | 0308 | | |
| 0109 | 200 | 0.140000 | 0309 | 208 | 0.221818 |
| 0110 | 360 | 0.069091 | 0310 | 428 | 0.072727 |
| 0111 | 668 | 0.160000 | 0311 | 400 | 0.203636 |
| 0112 | 600 | 0.101818 | 0312 | | |

Table 3
**Complexity of voting generated classifications in 4-th Duma (2004-2007)**

| 1 | 2 | 3 | 1 | 2 | 3 |
|---|---|---|---|---|---|
| 0401 | 101 | 0.360000 | 0601 | 168 | 0.216364 |
| 0402 | 220 | 0.101818 | 0602 | 204 | 0.289091 |
| 0403 | 270 | 0.141818 | 0603 | 256 | 0.265455 |
| 0404 | 295 | 0.101818 | 0604 | 255 | 0.147273 |
| 0405 | 249 | 0.325455 | 0605 | 179 | 0.194545 |
| 0406 | 385 | 0.143636 | 0606 | 365 | 0.085454 |
| 0407 | 378 | 0.372727 | 0607 | 260 | 0.221818 |
| 0408 | 268 | 0.303636 | 0608 | | |
| 0409 | 101 | 0.274545 | 0609 | 230 | 0.114545 |
| 0410 | 252 | 0.261818 | 0610 | 305 | 0.278182 |
| 0411 | 355 | 0.349091 | 0611 | 528 | 0.320000 |
| 0412 | 535 | 0.250909 | 0612 | 463 | 0.260000 |
| 0501 | 130 | 0.283636 | 0701 | 243 | 0.214545 |
| 0502 | 209 | 0.421818 | 0702 | 189 | 0.356364 |
| 0503 | 237 | 0.225455 | 0703 | 262 | 0.123636 |
| 0504 | 355 | 0.090909 | 0704 | 368 | 0.187273 |
| 0505 | 255 | 0.123636 | 0705 | 190 | 0.118182 |
| 0506 | 300 | 0.338182 | 0706 | 448 | 0.169091 |
| 0507 | 240 | 0.141818 | 0707 | 320 | 0.310909 |
| 0508 | | | 0708 | | |
| 0509 | 174 | 0.325455 | 0709 | 141 | 0.167273 |
| 0510 | 266 | 0.360000 | 0710 | 350 | 0.298182 |
| 0511 | 359 | 0.232727 | 0711 | 337 | 0.227273 |
| 0512 | 426 | 0.225455 | 0712 | | |

The numbers in the 3-rd column in Table 1 – 3, i.e. complexity of classifications based on voting results, demonstrate noticeable variability, though some trend are seen at once, by "unaided eye". Smoothed data, i.e. average value for half years, thereafter for years, and, finally, for whole period of every Duma activity, are presented in Table 4.

Table 4

**Smoothed complexity data**

|        | Half 1 | Half 2 | Half 3 | Half 4 | Half 5 | Half 6 | Half 7 | Half 8 |
|--------|--------|--------|--------|--------|--------|--------|--------|--------|
| Duma 2 | 0.711  | 0.448  | 0.387  | 0.251  | 0.432  | 0.394  | 0.340  | 0.290  |
| Duma 3 | 0.242  | 0.129  | 0.165  | 0.121  | 0.109  | 0.078  | 0.144  | 0.166  |
| Duma 4 | 0.196  | 0.302  | 0.247  | 0.257  | 0.199  | 0.239  | 0.195  | 0.251  |

|        | 1-st year | 2-nd year | 3-rd year | 4-th year |
|--------|-----------|-----------|-----------|-----------|
| Duma 2 | 0.606     | 0.332     | 0.415     | 0.320     |
| Duma 3 | 0.190     | 0.145     | 0.096     | 0.151     |
| Duma 4 | 0.249     | 0.252     | 0.217     | 0.217     |

| Duma 2 | Duma 3 | Duma 4 |
|--------|--------|--------|
| 0.418  | 0.147  | 0.235  |

It is curiously to compare the data presented in Table 4 with the averaged for every year stability index for the 3-rd Duma [Aleskerov et al, 2007]. These data, calculated using materials from the above cited book, are presented in Table 5.

Table 5

**Stability index in the 3-rd Duma**

| Year                                | 2000   | 2001   | 2002   | 2003   |
|-------------------------------------|--------|--------|--------|--------|
| Average stability index for one year | 0,5597 | 0,5627 | 0,5339 | 0,5090 |

Maximally possible value of stability index is equal to 1, minimally possible value is equal to 0. In contrast to the complexity data, which has a clear-cut minimum in 2002, stability index does not reach the maximum in this year. Perhaps it happens because stability indices were found basing on votes concerning only politically important issues, while in the present work all the votes are used. The used method itself was based on other reasoning, described in detail in the cited book.

It seems that low value of complexity in 2002 was due to creation of party "United Russia" and connected with this event attempts of straightening out the activity of Duma. It is surprising – at first sight – that in the 4-th Duma in the condition of constitutional majority of this party the level of complexity is noticeably higher than in the 3-rd Duma (0,235 opposite to 0,147), in which no party had majority.

One-month deputies classifications, found in order to filling Tables 1 – 3, let us to conduct a special investigation. An interest is attracted to correspondence between classes and deputies' fractions, dynamics of one-month classes changes, location of maximums and minimums and their connection with essential political events (such a connection was considered for one-month stability index in [Aleskerov et al, 2007]).

As it was marked above, value of complexity depends upon the parameters $k$ and $r$ of the essential algorithm. Let us consider this dependence in more detail. In order to do it, we calculated complexity for $k$ and $r$, changing from 5 to 10 inclusive. Tables 6, 7 and 8 contain values of complexity, calculated under parameters, changing within indicated limits, for 3 months: May, 1996; June, 2002, and February, 2005. These periods are related, correspondingly, to 2-nd, 3-rd and 4-th Duma; complexity has high (more 0.9), low (less 0.1) and middle (about 0.4) values.

Table 6

**Dependence of complexity on algorithm parameters for May, 1996**

| $r$ \ $k$ | 5 | 6 | 7 | 8 | 9 | 10 |
|---|---|---|---|---|---|---|
| 5 | 0.826667 | 0.876190 | 0.900000 | 0.933333 | 0.945455 | 0.922222 |
| 6 | 0.822222 | 0.873016 | 0.898810 | 0.921296 | 0.933333 | 0.945455 |
| 7 | 0.819048 | 0.870748 | 0.897959 | 0.920635 | 0.933333 | 0.942857 |
| 8 | 0.816667 | 0.869048 | 0.897321 | 0.920139 | 0.933333 | 0.940909 |
| 9 | 0.807407 | 0.862434 | 0.892857 | 0.916667 | 0.930864 | 0.939394 |
| 10 | 0.806667 | 0.861905 | 0.892857 | 0.913889 | 0.928889 | 0.938182 |

Table 7

**Dependence of complexity on algorithm parameters for June, 2002**

| $r$ \ $k$ | 5 | 6 | 7 | 8 | 9 | 10 |
|---|---|---|---|---|---|---|
| 5 | 0.120000 | 0.142857 | 0.114286 | 0.094444 | 0.093333 | 0.123636 |
| 6 | 0.100000 | 0.119048 | 0.095238 | 0.078704 | 0.085185 | 0.115152 |
| 7 | 0.085714 | 0.102041 | 0.081632 | 0.067460 | 0.076190 | 0.103896 |
| 8 | 0.075000 | 0.089286 | 0.071429 | 0.059028 | 0.066667 | 0.090909 |
| 9 | 0.066666 | 0.079365 | 0.063492 | 0.052469 | 0.059259 | 0.080808 |
| 10 | 0.060000 | 0.071429 | 0.057143 | 0.047222 | 0.053333 | 0.072727 |

Table 8
**Dependence of complexity on algorithm parameters for February 2005**

| r \ k | 5 | 6 | 7 | 8 | 9 | 10 |
|---|---|---|---|---|---|---|
| 5 | 0.360000 | 0.380952 | 0.414286 | 0.422222 | 0.440000 | 0.440000 |
| 6 | 0.344444 | 0.365079 | 0.398810 | 0.412037 | 0.437037 | 0.433333 |
| 7 | 0.333333 | 0.353741 | 0.382653 | 0.396825 | 0.415873 | 0.407792 |
| 8 | 0.333333 | 0.351190 | 0.379464 | 0.395833 | 0.416667 | 0.413636 |
| 9 | 0.318519 | 0.338624 | 0.365079 | 0.388889 | 0.409877 | 0.412121 |
| 10 | 0.320000 | 0.347619 | 0.371429 | 0.397222 | 0.415556 | 0.421818 |

In Tables 6 – 8 numbers in right bottom corner coincide with complexity values in the corresponding period. Convergence in every column is well appreciable that is completely naturally, because averaging is done over increasing number of runs of the same algorithm with the same initial data (neighborhood and dissimilarity matrices), differing only in random generator initiation. Numbers in rows slightly more variable, though in the considered limits any visible outliers are not presented. It is clear that in the 1-st and 3-rd cases minor variations of the chosen parameters results remain almost permanent ones. They can be used in order to achieve stable complexity values, satisfying practical needs. In the case of low complexity value (Table 7) it seems of expedient to increase parameter $k$ in 1-2 units to achieve a reasonable stability.

Generally it is reasonably to modify the suggested definition of complexity of AC problem through addition of adaptability in calculation of parameters $k$ and $r$, stopping by reaching a stable complexity value. It is supposed to consider this issue in the further investigations, though it should be mentioned that significant problems are not expected here.

The comparison with one other method of analysis of stability of political body was mentioned above in this section (see Table 5). It is supposed to consider these issues in more detail in a separate publication, especially concerning analysis of voting activity of political bodies, including RF Duma during its several convening.

Comparison of suggested method of solution of the general AC problem with other the most known approaches was done in the preprint [Rubchinsky, 2010]. Yet the comparison there was done only for model AC problems. It is interesting to compare results for considered in this section real data on voting in RF state Duma. Let us consider as an example the classification, based on voting in May, 2001 by one of the most known method – method of *K*-means. This method is described in book [Mirkin, 2010, p. 252] as follows.

"In general, the cluster finding process according to *K*-means starts from *K* tentative centroids and repeatedly applies two steps:
(a) collecting clusters around centroids,
(b) updating centroids as within cluster means,
– until convergence.
This makes much sense – whichever centroids are suggested first, as hypothetical cluster tendencies, they are checked then against real data and moved to the areas of higher density."

Assume the number of clusters is equal to 4. This number is determined by meaningful reasoning – in the 3-rd Duma 10 deputies' fractions and groups were presented, and therefore selection of 4 classes is expected. This does not mean that larger number of classes is impossible. The matter consists in the simple fact: union of several stable classes forms a stable class, too. At the same time the division into 4 classes is visible enough.

5 different classifications, found after 5 random determination of 4 initial centroids, are presented below. Remember (see Table 2) that objects are vectors with 248 components (value 1, −1 and 0). The first 4 numbers in every shown below classification are the initial centroids in *K*-means method. Every class is preceded by its cardinality. The number of steps is equal 1000, though convergence is reached after 200-300 steps.

**Classification 1**

195 465 459 202
 167
 3  4 14 17 21 22 23 26 27 28 30 40 44 45 48 53 55 56 58 61 64 67 69 70 74 77 78 79 82 85 90 91 92 93 94 100 102 110 111 114 116 121 122 123 124 126 128 129 134 136 137 139 145 147 148 155 157 162 167 168 174 175 179 180 182 183 188 195 197 199 200 202 203 206 208 209 213 214 216 218 221 223 225 229 232 236 238 239 240 241 247 249 251 253 257 259 262 266 271 272 273 274 275 276 278 281 282 285 286 292 294 298 300 301 307 311 316 320 321 324 326 331 338 339 340 343 344 347 358 361 366 370 372 373 380 381 386 388 393 394 397 398 400 404 405 407 410 413 415 417 418 420 422 424 428 431 434 438 440 441 444 445 448 450 452 455
 108
1  5 24 25 34 39 43 47 57 62 66 76 81 83 89 96 97 103 108 109 113 115

119 130 143 159 165 170 181 186 193 205 210 217 220 222 227 230 231 234 244
245 246 248 252 254 255 256 258 268 287 288 296 299 303 305 317 318 322 335
336 337 346 352 355 363 364 365 367 369 375 376 377 378 379 411 414 421 425
429 432 433 437 446 449 451 454 457 458 460 461 462 463 464 465 466 467 468
469 470 471 472 473 474 475 476 477
 79
  2  6  8 10 13 29 31 32 35 37 41 42 49 54 60 65 68 75 88 99 101 104 117
120 125 133 142 153 154 156 163 164 169 171 172 173 178 185 187 198 212 219
226 233 250 267 269 283 289 297 304 310 314 315 329 330 345 349 353 354 356
357 359 360 362 368 374 384 387 390 391 399 412 419 435 436 439 442
 125
  0  7  9 11 12 15 16 18 19 20 33 36 38 46 50 51 52 59 63 71 72 73 80
 84 86 87 95 98 105 106 107 112 118 127 131 132 135 138 140 141 144 146 149
150 151 152 158 160 161 166 176 177 184 189 190 191 192 194 196 201 204 207
211 215 224 228 235 237 242 243 260 261 263 264 265 270 277 279 280 284 290
291 293 295 302 306 308 309 312 313 319 323 325 327 328 332 333 334 341 342
348 350 351 371 382 383 385 389 392 395 396 401 402 403 406 408 409 416 423
426 427 430 443 447

**Classification 2**

199 221 141 440
 58
 14 17 25 39 43 62 66 76 81 83 116 123 136 145 170 181 193 199 205 206 213
223 227 241 244 245 247 249 254 255 256 258 276 294 296 305 317 318 322 336
363 369 372 375 378 379 413 414 418 425 428 429 433 434 438 451 452
 174
  3  4  5 21 22 23 24 26 27 28 30 40 44 45 47 48 53 55 56 57 58 61 64
 67 69 70 74 77 78 79 82 84 85 89 90 91 92 93 94 100 102 108 110 111 113
114 115 119 121 122 124 126 128 129 134 137 139 147 148 155 157 162 165 167
168 174 175 179 180 182 183 188 195 197 200 202 203 208 209 210 214 216 218
221 225 229 231 232 236 238 239 240 246 251 252 253 257 259 262 266 268 271
272 273 274 275 278 281 282 285 286 292 298 300 301 303 307 311 316 320 321
324 326 331 335 338 339 340 343 344 346 347 352 358 361 366 370 373 377 380
381 386 388
393 394 397 398 400 404 405 407 410 411 415 417 420 422 424 431 432
437 440 441 444 445 446 448 449 450 454 455 456 457
 127
  0  7  9 11 12 15 16 18 19 20 33 36 38 46 50 51 52 59 63 71 72 73 80
 86 87 95 98 105 106 107 112 118 127 130 131 132 135 138 140 141 144 146 149
150 151 152 158 160 161 166 176 177 184 189 190 191 192 194 196 201 204 207
211 215 224 228 234 235 237 242 243 260 261 263 264 265 270 277 279 280 284

290 291 293 295 299 302 306 308 309 312 313 319 323 325 327 328 332 333 334
341 342 348 350 351 371 382 383 385 389 392 395 396 401 402 403 406 408 409
416 423 426 427 430 443 447
 120
  1  2  6  8 10 13 29 31 32 34 35 37 41 42 49 54 60 65 68 75 88 96 97
 99 101 103 104 109 117 120 125 133 142 143 153 154 156 159 163 164 169 171 172
173 178 185 186 187 198 212 217 219 220 222 226 230 233 248 250 267 269 283
287 288 289 297 304 310 314 315 329 330 337 345 349 353 354 355 356 357 359
360 362 364 365 367 368 374 376 384 387 390 391 399 412 419 421 435 436 439
442 459 461 462 463 464 465 466 467 468 469 470 471 472 473 474 475 476 477

**Classification 3**

207  115  162  267
 127
  0  7  9 11 12 15 16 18 19 20 33 36 38 46 50 51 52 59 63 71 72 73 80
 84 86 87 95 98 105 106 107 112 118 127 130 131 132 135 138 140 141 144 146
149 150 151 152 158 160 161 166 176 177 184 189 190 191 192 194 196 201 204
207 211 215 224 228 234 235 237 242 243 260 261 263 264 265 270 277 279 280
284 290 291 293 295 302 306 308 309 312 313 319 323 325 327 328 332 333 334
341 342 348 350 351 371 382 383 385 389 392 395 396 401 402 403 406 408 409
416 423 426 427 430 443 447
 67
  1  5 34 57 89 96 97 103 108 113 115 119 143 159 186 210 217 220 222 227
230 231 244 246 248 252 268 287 288 299 303 318 335 337 346 352 355 364 365
367 376 377 421 432 437 446 454 457 460 461 462 463 464 465 466 467 468 469
470 471 472 473 474 475 476 477
 206
  3  4 14 17 21 22 23 24 25 26 27 28 30 39 40 43 44 45 47 48 53 55 56
 58 61 62 64 66 67 69 70 74 76 77 78 79 81 82 83 85 90 91 92 93 94 100
102 109 110 111 114 116 121 122 123 124 126 128 129 134 136 137 139 145 147
148 155 157 162 165 167 168 170 174 175 179 180 181 182 183 188 193 195 197
199 200 202 203 205 206 208 209 213 214 216 218 221 223 225 229 232 236 238
239 240 241 245 247 249 251 253 254 255 256 257 258 259 262 266 271 272 273
274 275 276 278 281 282 285 286 292 294 296 298 300 301 305 307 311 316 317
320 321 322 324 326 331 336 338 339 340 343 344 347 358 361 363 366 369 370
372 373 375 378 379 380 381 386 388 393 394 397 398 400 404 405 407 410 411
413 414 415 417 418 420 422 424 425 428 429 431 433 434 438 440 441 444 445
448 449 450 451 452 455 456
 79
  2  6  8 10 13 29 31 32 35 37 41 42 49 54 60 65 68 75 88 99 101 104 117
120 125 133 142 153 154 156 163 164 169 171 172 173 178 185 187 198 212 219
226 233 250 267 269 283 289 297 304 310 314 315 329 330 345 349 353 354 356

357 359 360 362 368 374 384 387 390 391 399 412 419 435 436 439 442

**Classification 4**

240  139   26 439
 150
   3   4  14  21  22  23  26  27  28  30  40  44  45  47  48  53  55  56  58  61  64  67  69
 70  74  77  78  79  82  85  90  91  92  93  94 100 102 109 111 114 121 122 124 126
128 129 134 137 139 147 148 155 157 162 167 168 174 175 179 180 182 183 188
195 197 199 200 202 203 208 209 213 214 216 218 221 225 229 232 236 238 239
240 251 253 257 259 262 266 271 272 273 274 275 276 278 281 282 285 286 292
298 300 301 307 311 316 320 321 324 326 331 338 339 340 343 344 347 358 361
366 370 373 380 381 386 388 393 394 397 398 400 404 405 407 410 415 417 420
422 424 431 440 441 444 445 448 450 455
 125
   0   7   9  11  12  15  16  18  19  20  33  36  38  46  50  51  52  59  63  71  72  73  80
 84  86  87  95  98 105 106 107 112 118 127 131 132 135 138 140 141 144 146 149
150 151 152 158 160 161 166 176 177 184 189 190 191 192 194 196 201 204 207
211 215 224 228 235 237 242 243 260 261 263 264 265 270 277 279 280 284 290
291 293 295 302 306 308 309 312 313 319 323 325 327 328 332 333 334 341 342
348 350 351 371 382 383 385 389 392 395 396 401 402 403 406 408 409 416 423
426 427 430 443 447
 125
   1   5  17  24  25  34  39  43  57  62  66  76  81  83  89  96  97 103 108 110
113 115 116 119 123 130 136 143 145 159 165 170 181 186 193 205 206 210 217
220 222 223 227 230 231 234 241 244 245 246 247 248 249 252 254 255 256 258
268 287 288 294 296 299 303 305 317 318 322 335 336 337 346 352 355 363 364
365 367 369 372 375 376 377 378 379 411 413 414 418 421 425 428 429 432 433
434 437 438 446 449 451 452 454 457 458 460 461 462 463 464 465 466 467 468
469 470 471 472 473 474 475 476 477
  79
   2   6   8  10  13  29  31  32  35  37  41  42  49  54  60  65  68  75  88  99 101 104 117
120 125 133 142 153 154 156 163 164 169 171 172 173 178 185 187 198 212 219
226 233 250 267 269 283 289 297 304 310 314 315 329 330 345 349 353 354 356
357 359 360 362 368 374 384 387 390 391 399 412 419 435 436 439 442

**Classification 5**

248   17 176 460
 120
   1   2   6   8  10  13  29  31  32  34  35  37  41  42  49  54  60  65  68  75  88  96  97
 99 101 103 104 109 117 120 125 133 142 143 153 154 156 159 163 164 169 171 172
173 178 185 186 187 198 212 217 219 220 222 226 230 233 248 250 267 269 283
287 288 289 297 304 310 314 315 329 330 337 345 349 353 354 355 356 357 359

360 362 364 365 367 368 374 376 384 387 390 391 399 412 419 421 435 436 439 442 459 461 462 463 464 465 466 467 468 469 470 471 472 473 474 475 476 477
 180
  3  4 14 17 21 22 23 25 26 30 40 44 45 53 58 62 64 66 67 69 74 76 78 79 85 89 90 91 92 94 102 108 111 113 114 115 116 119 122 123 124 128 136 137 139 145 147 148 162 165 167 168 170 175 180 181 182 188 195 197 199 205 206 208 213 214 218 221 223 227 229 231 232 236 238 239 240 244 246 247 249 251 252 253 255 256 257 262 266 268 271 272 273 274 275 276 278 282 285 286 292 294 298 300 301 303 307 311 316 317 318 320 321 322 324 326 331 336 338 339 340 343 344 346 347 352 358 361 363 366 369 370 372 373 375 377 378 379 380 381 386 388 393 394 397 398 400 404 405 407 410 411 413 414 415 417 418 420 422 424 425 428 429 431 433 434 437 438 440 441 445 446 448 450 452 454 455 456 457
 123
  0  7  9 11 12 15 18 19 20 33 36 38 46 50 51 52 59 63 71 72 73 80 86 87 95 98 105 106 107 112 118 127 130 131 132 135 138 140 141 144 146 149 150 151 152 158 160 161 166 176 177 184 189 190 191 192 194 196 201 204 207 211 215 224 228 234 235 237 242 243 260 261 263 264 265 270 277 279 280 284 290 291 293 295 299 302 306 308 309 312 313 319 323 325 327 328 332 333 334 341 342 348 350 351 371 382 383 389 392 395 396 401 402 403 406 408 409 416 423 426 427 430
 56
  5 16 24 27 28 39 43 47 48 55 56 57 61 70 77 81 82 83 84 93 100 110 121 126 129 134 155 157 174 179 183 193 200 202 203 209 210 216 225 241 245 254 258 259 281 296 305 335 385 432 443 444 447 449 451

   The shown classifications differ from one another considerably. There are no coinciding classifications. Write cardinality of classes for every classification in decreasing order:
   classification 1: 167  125  108  79;
   classification 2: 174  127  120  58;
   classification 3: 206  127   79  67;
   classification 4: 150  125  125  79;
   classification 5: 180  123  120  56.
Even classes, containing the same numbers of objects, for instance, 127 in 2-nd and 3-rd classifications, are coinciding not completely.
   Let us consider now 10 classifications, found by the suggested method for the same initial data, i.e. voting results. Among them there are 3 different classifications:

**Classification 1**

253
  1  3  4  5  21  22  23  24  25  26  27  28  30  34  39  40  43  44  45  47  48  53  55
 56  57  58  61  62  64  66  67  69  70  74  76  77  78  79  81  82  83  84  85  89  90  91
 92  93  94  96  97 100 102 103 108 109 110 111 113 114 115 119 121 122 124 126
128 129 134 137 139 143 147 148 155 157 159 162 165 167 168 170 174 175 179
180 181 182 183 186 188 193 195 197 199 200 202 203 205 208 209 210 214 216
217 218 220 221 222 225 227 229 230 231 232 236 238 239 240 245 246 248 251
252 253 254 255 256 257 258 259 262 266 268 271 272 273 274 275 278 281 282
285 286 287 288 292 296 298 299 300 301 303 305 307 311 316 317 318 319 320
321 322 324 326 331 335 336 337 338 339 340 343 344 346 347 352 355 358 361
363 364 365 366 367 369 370 373 375 376 377 378 379 380 381 386 388 393 394
397 398 400 404 405 407 410 411 414 415 417 420 421 422 424 425 429 431 432
433 437 440 441 444 445 446 448 449 450 451 454 455 456 457 458 460 461 462
463 464 465 466 467 468 469 470 471 472 473 474 475 476 477 478
 79
  2  6  8 10 13 29 31 32 35 37 41 42 49 54 60 65 68 75 88 99 101 104 117
120 125 133 142 153 154 156 163 164 169 171 172 173 178 185 187 198 212 219
226 233 250 267 269 283 289 297 304 310 314 315 329 330 345 349 353 354 356
357 359 360 362 368 374 384 387 390 391 399 412 419 435 436 439 442 459
 125
  0  7  9 11 12 15 16 18 19 20 33 36 38 46 50 51 52 59 63 71 72 73 80
 86 87 95 98 105 106 107 112 118 127 130 131 132 135 138 140 141 144 146 149
150 151 152 158 160 161 166 176 177 184 189 190 191 192 194 196 201 204 207
211 215 224 228 234 235 237 242 243 260 261 263 264 265 270 277 279 280 284
290 291 293 295 302 306 308 309 312 313 323 325 327 328 332 333 334 341 342
348 350 351 371 382 383 385 389 392 395 396 401 402 403 406 408 409 416 423
426 427 430 443 447 453
 22
 14 17 116 123 136 145 206 213 223 241 244 247 249 276 294 372 413 418 428 434
438 452

**Classification 2**

221
  1  3  4  5 21 22 23 24 26 27 28 30 34 40 44 45 47 48 53 55 56 57 58
 61 64 67 69 70 74 77 78 79 82 84 85 89 90 91 92 93 94 96 97 100 102
103 108 109 110 111 113 114 115 119 121 122 124 126 128 129 134 137 139 143
147 148 155 157 159 162 165 167 168 174 175 179 180 181 182 183 186 188 195
197 199 200 202 203 208 209 210 214 216 217 218 220 221 222 225 227 229 230
231 232 236 238 239 240 246 248 251 252 253 257 259 262 266 268 271 272 273
274 275 278 281 282 285 286 287 288 292 298 299 300 301 303 307 311 316 319
320 321 324 326 331 335 337 338 339 340 343 344 346 347 352 355 358 361 364

365 366 367 370 373 376 377 380 381 386 388 393 394 397 398 400 404 405 407
410 411 415 417 420 421 422 424 431 432 437 440 441 444 445 446 448 449 450
454 455 456 457 458 460 461 462 463 464 465 466 467 468 469 470 471 472 473
474 475 476 477 478
 79
  2  6  8 10 13 29 31 32 35 37 41 42 49 54 60 65 68 75 88 99 101 104 117
120 125 133 142 153 154 156 163 164 169 171 172 173 178 185 187 198 212 219
226 233 250 267 269 283 289 297 304 310 314 315 329 330 345 349 353 354 356
357 359 360 362 368 374 384 387 390 391 399 412 419 435 436 439 442 459
 125
  0  7  9 11 12 15 16 18 19 20 33 36 38 46 50 51 52 59 63 71 72 73 80
 86 87 95 98 105 106 107 112 118 127 130 131 132 135 138 140 141 144 146 149
150 151 152 158 160 161 166 176 177 184 189 190 191 192 194 196 201 204 207
211 215 224 228 234 235 237 242 243 260 261 263 264 265 270 277 279 280 284
290 291 293 295 302 306 308 309 312 313 323 325 327 328 332 333 334 341 342
348 350 351 371 382 383 385 389 392 395 396 401 402 403 406 408 409 416 423
426 427 430 443 447 453
 54
 14 17 25 39 43 62 66 76 81 83 116 123 136 145 170 193 205 206 213 223 241
244 245 247 249 254 255 256 258 276 294 296 305 317 318 322 336 363 369 372
375 378 379 413 414 418 425 428 429 433 434 438 451 452

**Classification 3**

243
  1  3  4  5 14 17 21 22 23 24 26 27 28 30 34 40 44 45 47 48 53 55 56
 57 58 61 64 67 69 70 74 77 78 79 82 84 85 89 90 91 92 93 94 96 97 100
102 103 108 109 110 111 113 114 115 116 119 121 122 123 124 126 128 129 134
136 137 139 143 145 147 148 155 157 159 162 165 167 168 174 175 179 180 181
182 183 186 188 195 197 199 200 202 203 206 208 209 210 213 214 216 217 218
220 221 222 223 225 227 229 230 231 232 236 238 239 240 241 244 246 247 248
249 251 252 253 257 259 262 266 268 271 272 273 274 275 276 278 281 282 285
286 287 288 292 294 298 299 300 301 303 307 311 316 319 320 321 324 326 331
335 337 338 339 340 343 344 346 347 352 355 358 361 364 365 366 367 370 372
373 376 377 380 381 386 388 393 394 397 398 400 404 405 407 410 411 413 415
417 418 420 421 422 424 428 431 432 434 437 438 440 441 444 445 446 448 449
450 452 454 455 456 457 458 460 461 462 463 464 465 466 467 468 469 470 471
472 473 474 475 476 477 478
 79
  2  6  8 10 13 29 31 32 35 37 41 42 49 54 60 65 68 75 88 99 101 104 117
120 125 133 142 153 154 156 163 164 169 171 172 173 178 185 187 198 212 219
226 233 250 267 269 283 289 297 304 310 314 315 329 330 345 349 353 354 356
357 359 360 362 368 374 384 387 390 391 399 412 419 435 436 439 442 459

125
0 7 9 11 12 15 16 18 19 20 33 36 38 46 50 51 52 59 63 71 72 73 80
86 87 95 98 105 106 107 112 118 127 130 131 132 135 138 140 141 144 146 149
150 151 152 158 160 161 166 176 177 184 189 190 191 192 194 196 201 204 207
211 215 224 228 234 235 237 242 243 260 261 263 264 265 270 277 279 280 284
290 291 293 295 302 306 308 309 312 313 323 325 327 328 332 333 334 341 342
348 350 351 371 382 383 385 389 392 395 396 401 402 403 406 408 409 416 423
426 427 430 443 447 453
32
25 39 43 62 66 76 81 83 170 193 205 245 254 255 256 258 296 305 317 318
322 336 363 369 375 378 379 414 425 429 433 451

Classifications 1 и 2 are encountered 4 times from 10, classification 3 – 2 times. Write cardinality of classes for every classification in decreasing order:

classification 1: 253 125 79 22;
classification 2: 221 125 79 54;
classification 3: 243 125 79 32.

Note that some classes are found by both methods. Yet the stability of classifications, found by the suggested algorithm, significantly exceeds the stability of classifications, found by *K*-means methods, for the same AC problem. This directly noticeable fact is established exactly by the algorithm of stability calculation of a classification family, considered at the end of subsection 3.4. Pay attention to practical absence (in 4 cases among 5 ones, found by *K*-means method) of classes, containing more than 200 objects, while such classes are present in all the classifications found by the suggested algorithm. The essence of the matter if not is proved but is illustrated by the example from preprint [Rubchinsky, 2010], where the application of *K*-means method for the set of points, shown in Fig. 8, is presented. It is clear that this method cannot work in similar situations, which cannot be a priori excluded in real AC problems. At the same time in the construction of classifications based on voting results, such non-uniform case are occurred frequently enough.

Instability of classifications, found by *K*-means method, is detected as well in the other checked periods, particularly, in February, 2005 and April, 1996.

In order to resume the present section it should be remarked the following. The notion of complexity is relevant to arbitrary AC problems, whose solutions are considered as partitions of the initial set of objects. This

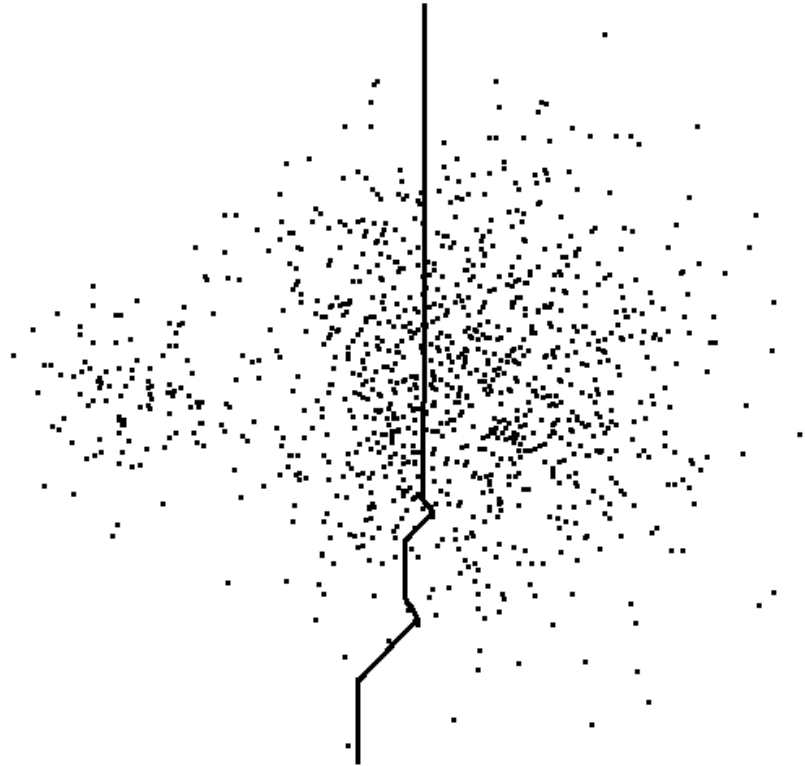
Fig. 8. *K*-means method for the model example

notion itself does not bear any meaningful load. Some AC problems from preprint [Rubchinsky, 2010] are complex ones in the above sense, despite their solutions are intuitively obvious. In examples 10 and 11 from the same preprint solutions are not obvious ones, while their formal complexity is low enough. Yet for some types of AC problems the notion of complexity can acquire a special interpretation. For analysis of political bodies making collective decisions by voting, the complexity corresponds to inconsistency, incoordination, irrationality of politics – independently of presence or absence of majority of some deputies' group, even if all the members of every fraction vote similarly. For "tossing" deputies or / and whole fractions the corresponding classes become poorly distinguished and partially perplexing that results in relatively high value of complexity of their classifications.

Thus, the complexity is not defined by the results of separate votes, but rather by the set of all such results. The situation slightly likes the definition of choice functions, whose properties are not determined by the separate results of variants choice, but rather by interrelations between choices from various presented subsets of a given general set.

### 6. Conclusion

The main goal of the presented work consists in introducing of new notion of AC problem complexity and to its use in analyzing political processes. Many important issues concerning AC problems are not considered in the presented work – first of all from lack of the room as well as my disinclination to overloading the exposition. It is supposed to consider these issues in next publications. Some topics were mentioned above in subsection 3.4. The other ones are briefly mentioned below.

1. It is supposed to analyze voting results basing on AC in more detail, in RF Duma as well as in the other political bodies, in a special publication.

2. It is supposed to apply the suggested approach to stock market analysis, considering changes of complexity of constructed classifications in an attempt to predict some events.

3. Determination – even if the experimental one – of stochastic characteristics of considered classifications can allow us to obtain more exact and reliable estimations of the considered indices. It is supposed to be done in further investigations.

4. It is desirable to elaborate an adaptive modification of the suggested AC algorithm for determination its essential parameters $k$ and $r$. In particular, complexity calculation can be accomplished under different values of these parameters even for AC problem from the same family (for instance, analyzing voting results in one body in different periods).

5. Informal character of AC problems requires design of a special interactive computer system, as it was mentioned in subsection 3.4. In the framework of such a system it will be possible to change algorithm details, visually present results, and, finally to make final choice of classifications.

The author is grateful to F.T. Aleskerov for his support and attention to this work, B.G. Mirkin for attentive and benevolent reviewing, which helped to improve the quality and style of the material exposition, V.I. Jakuba for help in program realizations and N.Y. Blagoveschenskiy for helpful discussions.

The article was prepared within the framework of the Basic Research Program at the National Research University Higher School of Economics (HSE) and supported within the framework of a subsidy granted to the HSE by the Government of the Russian Federation for the implementation of the Global Competitiveness Program.